\documentstyle[psfig,wrapfig]{IOPConf}    
\begin{document}
\frenchspacing

\title{Structure Effects on Coulomb Dissociation of $^8$B}

\longauthor{F.M. Nunes$^{1)}$, R. Shyam$^{2)}$ and I.J. Thompson$^{3)}$}
{F.M. Nunes}

\address{
$^{1)}$ Department of Physics, Instituto Superior Tecnico, Lisbon, Portugal\\ $^{2)}$ Saha Institute for Nuclear Physics, Calcutta, India \\ 
$^{3)}$ Department of Physics, University of Surrey, Guildford, UK}

\beginabstract
Coulomb Dissociation provides an alternative method for determining the
radiative capture cross sections at astrophysically relevant low
relative energies.  For the breakup of $^8$B on $^{58}$Ni, we calculate
the total Coulomb Dissociation cross section and the angular
distribution for E1, E2 and M1.  Our calculations are performed first within
the standard first order semiclassical theory of Coulomb
Excitation, including the correct three body kinematics, and later
including the projectile-target nuclear interactions.  

We study the dependence of the Coulomb Dissociation cross section on
the structure models assumed for the projectile.  A range of  potential
models for $^7$Be+p are compared:  we look at the effect of potential shapes,
deformation and inclusion of inelastic channels in the
projectile states.  
We analyse the relative E1
and E2 components and investigate the relation between the measured
cross section and the S-factor $S_{17}$.  
Preliminary Coulomb-nuclear interference results are presented.
\endabstract


\section{Introduction}
Coulomb Dissociation (CD) experiments have brought new insight to
the low energy capture reaction cross section $^7$Be(p,$\gamma$)$^8$B
which is a crucial ingredient for the solar neutrino puzzle \cite{neut}.
The main idea resides on the fact that in suitable conditions,
solely the Coulomb field is
responsible for the breakup of the projectile, and nuclear uncertainties
in the projectile-target interaction do not affect the total cross section. 
On the other hand, all CD measurement are {\it contaminated} by 
a non-negligible E2 component and although
the astrophysical S-factor $S_{17}$ (related to the $^7$Be proton capture
 rate  at $E= 20$ KeV) is determined by the E1 component only, to extract information 
from the CD data requires information on the E2 contribution.
In this work we study the dependence of the total CD cross section on the structure
models assumed for the projectile. 

The experimental state of the art is very encouraging.
The first Coulomb Dissociation measurements were performed
at RIKEN \cite{moto1} using a heavy target, a $^8$B beam of
$46.5$ MeV/A and complete kinematics.
There have also been some measurements on a 
lighter target at Notre Dame \cite{nd}  with a $^8$B beam of 3.225 MeV/A
where only one of the fragments was detected. 
Recently the experiment at RIKEN was repeated with
better statistics  and a wider angular range in order to obtain more
information on the E2 contribution.
Coulomb Dissociation of $^8$B being measured at 
relativistic energies in GSI \cite{gsi} will soon be available.

Considerable theoretical effort has been offered to clarifying
the relation between the Coulomb Dissociation process and the
direct capture reaction. It is generally believed that as long as the
minimum impact parameter is larger than the sum of the two nuclei radii
plus a few fm,
the nuclear effects as well as nuclear-Coulomb interference can be neglected.
This appears to be the case \cite{shyam1} for the RIKEN experiment.
Higher order terms have been discussed, namely
it has been suggested that 
E1-E2 interference can partially cancel the E2 contribution \cite{esb}.
For the purpose of this work, we will consider only E1, E2 and M1
contributions to first order, and including proper three body
kinematics\cite{shyam2}, and then explore in a preliminary way the true effects
of nuclear excitations.

\section{Virtual Photon Number}

\begin{wrapfigure}{R}{.60\textwidth}
\psfig{figure=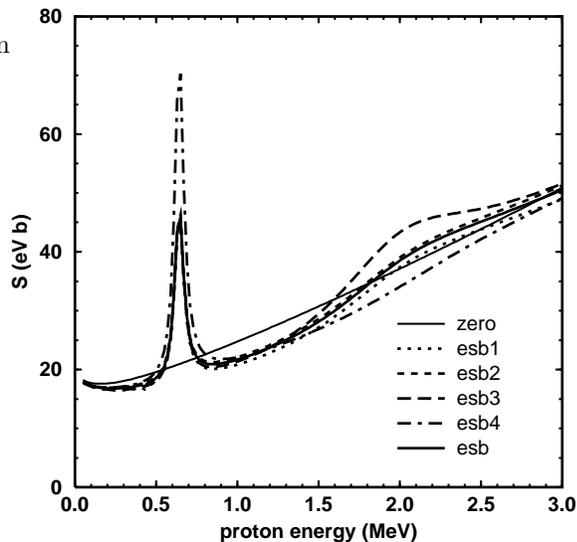,width=.60\textwidth}
	\caption{\captionfont The effect on the proton capture S-factor of variations
	in the nuclear interaction for the continuum $^8$B states.}
	\label{fig:sfacesb}
\end{wrapfigure}
Assuming that the nuclear interaction of projectile-target is weak and
that post-acceleration effects are small, the Coulomb Dissociation process
can be treated semi-classically \cite{alder} in which the projectile-target
relative motion takes place on a Rutherford trajectory. 
In first order, the differential cross section is the sum of the electromagnetic multipole
components and  can written as a product of a virtual photon
number VPN (depending only on the kinematics of the projectile-target relative motion
and the excitation energy), and a photo-disintegration cross section of
the projectile (defining the internal structure of the projectile):  
\begin{equation}
\frac{d \sigma^{CD}_{E \lambda}}{d \Omega \; dE_{\gamma}} = \frac{1}{E_{\gamma}} \: 
\frac{d n_{E \lambda}}{d \Omega} \: \sigma^{photo}_{E \lambda}
\end{equation}
The differential virtual photon number
(as a function of the relative energy of the projectile
fragments and the scattering angle) determines which parts of the 
energy spectrum of the projectile
are relevant for a particular set-up of a Coulomb dissociation process. 
The lower part of the projectile
spectrum is enhanced for  the lower beam energies and the smaller scattering angles. 
In the low-energy experiment of \cite{nd}, where $E_{beam}=3.225$ MeV/A 
and $\theta\simeq45 ^{\circ}$, the E1 VPN is a slowly decreasing function of $E_{rel}$,
so we expect the CD cross section will be sensitive to
variations in the negative parity continuum of $^8$B. On the other hand, the E2 VPN decreases steeply thus the CD cross section will be mainly sensitive to the lower energy part of the positive parity continuum of $^8$B.


\section{Structure Models for $^8$B}

The breakup of $^8$B proceeds through E1 into $1^-,2^-,3^-$ and 
through E2 (and M1) into $0^+,1^+,2^+,3^+$.
The ground state (g.s.) of $^8$B ($2^+$) can be qualitatively described 
as $p+^7$Be  in a $p_{3/2}$ relative motion. There are no resonances 
in the continuum for the negative parity states up to 3 MeV, which leaves some 
freedom for the parameter choices in potential models \cite{kim}.
There are two low lying resonances: 
a $1^+$  state at  $E=0.63$ MeV that is quite narrow,
and a broader  $3^+$ state at $E=2.18$ MeV. 
The constraint imposed by these two resonances accepts a wide
range of potential models. 

\subsection{Uncertainties in the continuum of $^8$B}

\begin{wrapfigure}{R}{.60\textwidth}
\psfig{figure=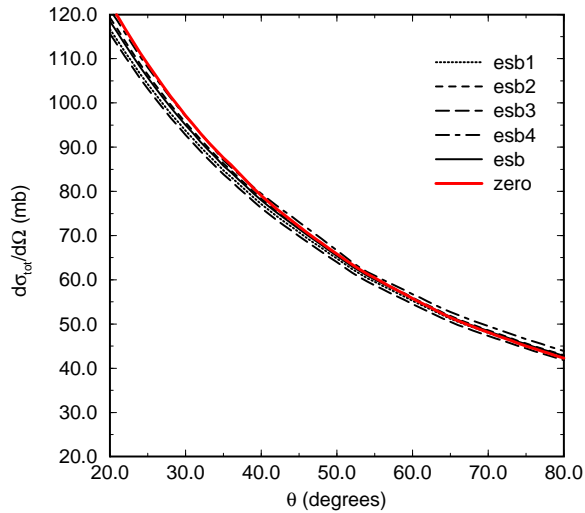,width=.60\textwidth}
	\caption{\captionfont Effects of continuum nuclear potentials on the CD cross section
	at 3.225 MeV/A.}
	\label{fig:dc2esb}
\end{wrapfigure}

First we evaluate the uncertainty associated with the nuclear interaction
in the continuum, which is mainly related to the negative parity states.
For this we take the model from \cite{esb} [esb] and modify by $\simeq20\%$ the strength
of the nuclear interaction for all but the $2^+,1^+, 3^+$ states [esb1,esb2,esb3,esb4].
In fig. \ref{fig:sfacesb} we show the resulting S-factor as a function
of the proton CM energy. In fig. \ref{fig:dc2esb} we show the differential
cross section for the Coulomb dissociation on a $^{58}$Ni target for
$E_{beam}=3.225$ MeV/A (the Notre Dame experimental conditions \cite{nd}).
The Coulomb dissociation calculations use the method of \cite{alder} with
proper three-body kinematics \cite{shyam2}.
In these figures  we also include the results for the extreme case where no
nuclear interaction is included in the $^7$Be-p continuum [zero].
	
\begin{wrapfigure}{R}{.60\textwidth}
	\psfig{figure=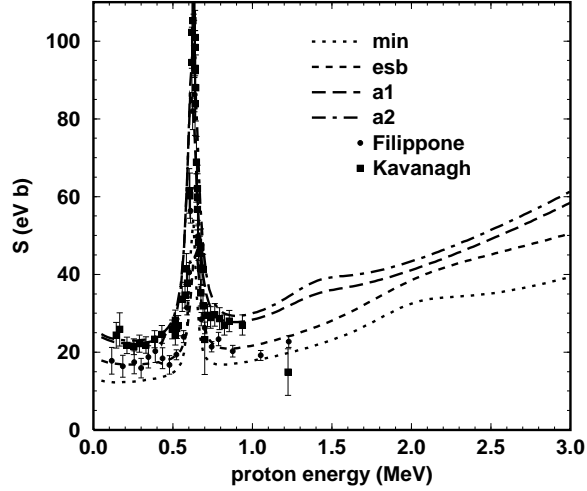,width=.60\textwidth}
	\caption{\captionfont The proton capture S-factor for a set of  structure models for $^8$B
	(data from \protect\cite{fil83,kav69} with the updated normalisations).}
	\label{fig:sfacall}
\end{wrapfigure}

The very low energy S-factor is not affected by these small variations
on parameters for the continuum interactions (for the
set of models under consideration S(E=50 keV) differ by less than $1\%$)
however there are considerable differences in the
S-factor curves above $0.4$ MeV. We find that 
the breakup cross section at low energies is not strongly
influenced by these parameter variations.  
Thus the uncertainties in the nuclear interactions for
the $^8$B scattering states have a negligible influence on the Coulomb Dissociation
differential cross section for the Notre Dame set-up. This conclusion holds for
lower and higher beam energies.



\subsection{Shape parameters - Overall normalisation}

\begin{wrapfigure}{r}{.60\textwidth}
	\psfig{figure=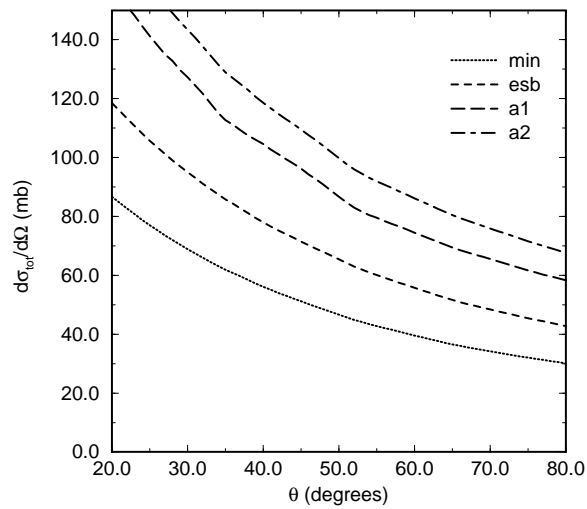,width=.60\textwidth}
	\caption{\captionfont The  dependence of the differential Coulomb dissociation cross section for a set of  structure models for $^8$B.}
	\label{fig:dc2all}
\end{wrapfigure}

In this section we compare the results obtained with different
potential models. All potential models reproduce the binding energy
and the energy of the two resonances. We choose four models from the literature
based on a Woods-Saxon shape:
{\bf [esb]} \cite{esb} with  $R=2.39$ fm  and $a=0.52$ fm; 
{\bf [kim]} \cite{kim} with $R=0.52$ fm and $a=0.52$ fm; 
{\bf [a1]} and {\bf [a2]} with the same shape parameters as {\bf [kim]} 
but including deformation
of the $^7$Be nuclear and reorientation effects \cite{nunes1}; 
{\bf [c1]} and {\bf [c2]}
that consist on the extension of {\bf [a1]} and {\bf [a2]} including the inelastic channels
corresponding to all excited states
of the $^7$Be core \cite{nunes1}.
We also include the predictions for what we call the minimum model {\bf [min]}
where we take low values for both radius and diffuseness parameters verging
on the unrealistic ($R=2.0$ fm and $a=0.4$ fm). 

\begin{wrapfigure}{R}{.50\textwidth}
	\psfig{figure=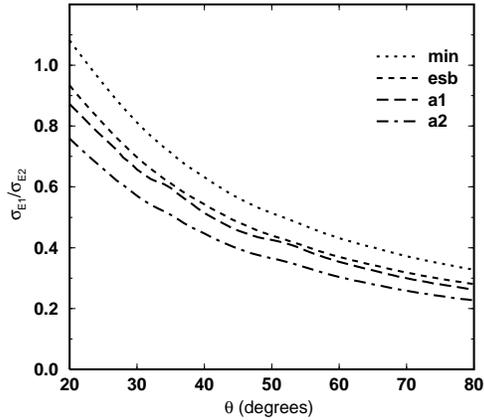,width=.50\textwidth}
	\caption{\captionfont The E1 to E2 ratio in the Coulomb dissociation for a set of
structure models of $^8$B incident at 3.225 MeV/A on $^{58}$Ni.}
	\label{fig:e1e2all}
\end{wrapfigure}

The calculated S-factors for this set of models are presented in fig. \ref{fig:sfacall}.
It is now well accepted that all potential models predict roughly the
same low  energy behaviour for the S-factor apart from an overall normalisation 
\cite{bar,nunes1}.
This same effect is seen in the differential cross section for
Coulomb dissociation and  is clearly shown in fig. \ref{fig:dc2all} where we plot the results
for $^8$B Coulomb dissociation on a $^{58}$Ni target for $E_{beam}=3.225$ MeV/A
for some of the models. We find that the CD cross section is not very sensitive to 
the differences in the $^8$B spectrum above $E_{rel}\sim 0.5$ MeV.
In fig. \ref{fig:e1e2all} 
we show the ratio between E1 and E2 contributions
to the differential cross section for some of the models. We find that the 
curves differ by a constant factor, and that
the models that have a larger E1 fraction predict a smaller total cross section.

\begin{wrapfigure}{r}{.50\textwidth}
	\psfig{figure=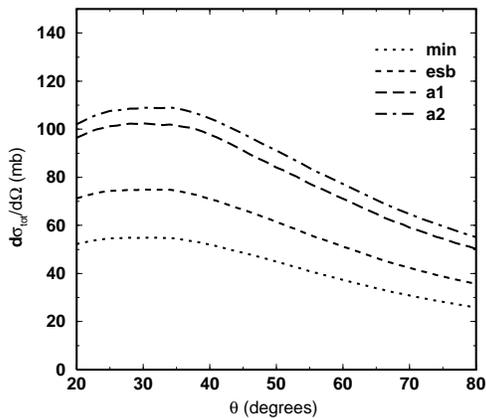,width=.50\textwidth}
	\caption{\captionfont The  dependence of the Coulomb dissociation cross section on 
$^{197}$Au on the  structure model for $^8$B.}
	\label{fig:dc2au197}
\end{wrapfigure}

At Notre Dame it was not possible to measure the differential cross
section. Instead the data refers to a total cross section integrated
from $\theta=39^{\circ}-51^{\circ}$.
In table \ref{ndtb} the results obtained for the total cross section 
predicted for the Notre Dame experiment are shown. 
As in \cite{shyam2}, the predictions are much larger than the data.
We confirm this result, and that it is not possible to 
reproduce the measured result for any of our models of $^8$B structure.  
One could at first think that the E2 contribution  was being over estimated,
but in some cases even just the E1 cross section is too large (see table \ref{ndtb}).

\begin{table}[t]
\centering
 \begin{tabular}{||l|r|r|r|r||} \hline   
 Model 
& $\frac{\sigma (tot)}{\sigma (ruth)} $ 
& $\frac{\sigma (E1)}{\sigma (ruth)}$ 
& $\frac{\sigma (E2)}{\sigma  (E1)}$ 
&  S(50 keV) (eV b)\\
  \hline\hline
min  &   $2.07 \times 10^{-2}$ &  $0.75 \times 10^{-2}$ & $1.74$ & $12.7$  \\ \hline
esb  &   $2.89 \times 10^{-2}$ &  $0.95 \times 10^{-2}$ & $2.04$ & $17.9$  \\ \hline
kim  &  $3.88 \times 10^{-2}$ &  $1.16 \times 10^{-2}$ & $2.35$ &  $22.5$ \\ \hline
a1    &  $3.86 \times 10^{-2}$ &  $1.23 \times 10^{-2}$ & $2.14$ &  $23.9$ \\ \hline
a2    &  $4.41 \times 10^{-2}$ &  $1.27 \times 10^{-2}$ & $2.48$ &  $24.7$ \\ \hline
c1    &  $3.39 \times 10^{-2}$ &  $1.06 \times 10^{-2}$ & $2.20$ &  $20.9$ \\ \hline
c2    &  $4.04 \times 10^{-2}$ &  $1.17 \times 10^{-2}$ & $2.46$ &  $22.8$ \\ \hline
exp              &   $(0.68 - 1.09)\times 10^{-2}$ &    &     &          \\\hline \hline
\end{tabular}
\caption{\captionfont The results obtained for the total Coulomb Dissociation cross
section of $^8$B on $^{58}$Ni, for a set of $^8$B models.}
\label{ndtb}
\end{table}

This great mismatch with the low-energy data  may signify 
that the $^7$Be-p component of the $^8$B states is 
very small and that other cluster components are more important. This
is unlikely since the models are able to reproduce the capture
data within a reasonable accuracy. If that were the case then one
 would be in disagreement with microscopic predictions \cite{brown}.
The mismatch with the data can  alternatively mean that 
there is something missing in the reaction
theory used to describe the process: higher order contributions
inducing E1-E2 or Coulomb-nuclear interference may play a  more relevant role
than what was initially  thought.

From our predictions it is clear that even if the models were in agreement 
with the experimental result it would still be hard to 
distinguish between models since some models predict similar cross
sections but have different E1 and E2 components and thus predict quite
different low energy S-factor for the capture process (dominated by E1 only). 
From this data
 there is no indication as to
the relative amount of E1 and E2. 
Thus it may be helpful to perform measurement at different beam energies,
different angular ranges, or using different targets in order to get
more information on the relative amounts of E1 and E2 contributions.

We also show (fig. \ref{fig:dc2au197}) the resulting differential cross section
if the $^{58}$Ni target where replaced by a $^{197}$Au target. This is one of the
experiments that will be run in a few months \cite{kolata}.
 
\section{Improving the reaction theory}

\begin{wrapfigure}{r}{.60\textwidth}
	\psfig{figure=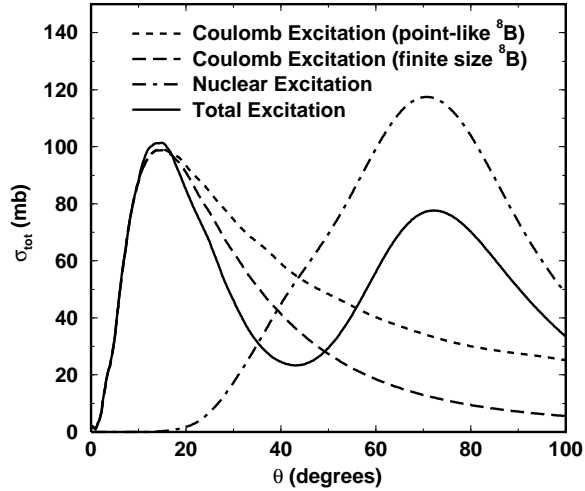,width=.60\textwidth}
	\caption{\captionfont The quantum mechanical calculations for the differential Coulomb dissociation cross section using [esb].}
	\label{fig:qm1esb}
\end{wrapfigure}

\begin{wrapfigure}{R}{.60\textwidth}
	\psfig{figure=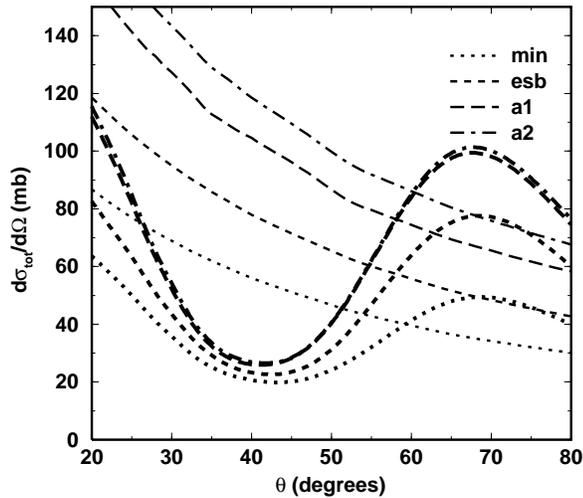,width=.60\textwidth}
	\caption{\captionfont The  $^8$B model dependence of the
differential cross section within a 1 step quantum calculation (thick  lines) compared with the semi-classical calculations.}
	\label{fig:qm1all}
\end{wrapfigure}

Given the difficulties that arose in trying to understand the existing Notre
Dame data we decided to improve the reaction theory by taking nuclear-excitation effects into account.
We performed prior-DWBA breakup calculations by discretising the continuum,
including $s_{1/2}$, $p_{1/2}$, $p_{3/2}$, $d_{3/2}$, $d_{5/2}$ partial waves
up to E(p-$^7$Be)$=3$ MeV. 
The results shown in fig. \ref{fig:qm1esb} use 13 bins ($9$ bins of $100$ keV centred at 
$0.15; 0.25; ...; 0.95$ MeV and $4$ bins of $500$ keV centred
at 1.25; 1.75; 2.25;2.75) , and in fig. \ref{fig:qm1all} use 6 bins 
($4$ bins of $250$ keV centred at 
0.125; 0.375; 0.625; 0.0.875 MeV and 2 bins of $1000$ keV centred
at 1.5; 2.5) and use $l_{max}=600 \hbar$ and $R_{max}=300$ fm for
the T-matrix integrals. 

The E1 and E2 multipoles give cross sections 
(long-dashes in fig. \ref{fig:qm1esb}) which reproduce the previous 
semiclassical results at forward angles. They agree at all angles 
(short-dashes in fig. \ref{fig:qm1esb}) if the $r^{-\lambda -1}$  multipoles
are extrapolated to small radii. This point-projectile approximation 
(made in the semiclassical method) becomes invalid for scattering angles
larger than $20 ^{\circ}$. However, at larger angles and smaller impact parameters,
when the g.s. wavefunction of the projectile overlaps with the 
target interior, nuclear effects become simultaneously important.

Using a Becchetti-Greenlees \cite{becc} proton potential, and a $^7$Li potential
\cite{moroz} for $^7$Be, the pure nuclear breakup 
(dot-dashed in fig. \ref{fig:qm1esb}) becomes significant beyond $25 ^{\circ}$.
When nuclear and Coulomb multipoles are included coherently, there are already
small effects below $20 ^{\circ}$, a large Coulomb-nuclear interference minimum between $25 ^{\circ}$ and $50 ^{\circ}$, and a nuclear-dominated
peak at $75 ^{\circ}$ (solid-line in fig. \ref{fig:qm1esb}).
This large nuclear effect is present even though the elastic Coulomb
+ nuclear cross section only drops to $90\%$ of the Rutherford cross
section at $70 ^{\circ}$, because of the large halo-like
size of proton wavefunction in the g.s. of $^8$B.

Fig. \ref{fig:qm1all} shows the effects of the $^8$B structure on the
pure-Coulomb and on the DWBA nuclear+Coulomb predictions.
The dependencies on structure are largely similar in both cases,
although the differences between the deformed-core models (a1 and a2)
is reduced in the DWBA predictions. 
In all $^8$B models the interference minimum between $30 ^{\circ}$ and $50 ^{\circ}$
is present, and we begin to understand the low cross sections measured 
between $31 ^{\circ}$ and $59 ^{\circ}$ in the ND experiment. 
The horizontal-axis on both figures \ref{fig:qm1esb} and \ref{fig:qm1all}
refers to the scattering angle $\theta$($^8$B$^*$).

\ack
UK support from the EPSRC grants GR/J/95867
and Portuguese support from JNICT PRAXIS/PCEX/P/FIS/4/96 are acknowledged

\references

\list
 {[\arabic{enumi}]}{\settowidth\labelwidth{[99]}\leftmargin\labelwidth
 \advance\leftmargin\labelsep
 \usecounter{enumi}}
 \def\newblock{\hskip .11em plus .33em minus .07em}
 \sloppy\clubpenalty4000\widowpenalty4000
 \sfcode`\.=1000\relax
 \let\endthebibliography=\endlist

\itemsep=-1pt
\bibitem{neut} J.N. Bahcall, {\em Neutrino Astrophysics} (Cambridge University Press, New York, 1989)
\bibitem{moto1} T. Motobayashi et al., Phys. Rev. Letts. {\bf 73} (1994) 2680
\bibitem{nd} Johannes von Schwarzenberg et al., Phys. Rev. {\bf C53} (1996) R2598
\bibitem{moto2} T. Kikuchi et al., Phys. Letts. {\bf B391} (1997) 261
\bibitem{gsi} Klaus Suemmerer, {\em private communication},  December 1997
\bibitem{shyam1} R. Shyam, I.J. Thompson and A.K. Dutt-Mazumder, Phys. Letts. {\bf B371} (1996) 1
\bibitem{esb} H. Esbensen and G. Bertsch, Nucl. Phys. {\bf  A600} (1996) 37
\bibitem{shyam2} R. Shyam and I.J. Thompson, Phys. Letts. {\bf B415} (1997) 315
\bibitem{alder} K. Alder and A. Winther, {\em Electromagnetic Excitation},
(North Holland, Amsterdam, 1975)
\bibitem{bar} F.C. Barker, Aust. J. Phys. {\bf 33} (1980) 177
\bibitem{kim} K.H. Kim, M.H. Park and B.T. Kim, Phys. Rev. {\bf C53} (1987) 363
\bibitem{nunes1} F. Nunes, R. Crespo and I.J. Thompson, Nucl. Phys. {\bf A615} (1997) 69,
erratum {\bf A627} (1997) 747, nucl-th/9709070
\bibitem{brown} B.A. Brown, A. Cs\'ot\'o and R. Sherr, Nucl. Phys. {\bf A597} (1996) 66
\bibitem{kolata} J.J. Kolata,  {\em private communication}, December 1997
\bibitem{moroz} Z. Moroz et al., Nucl. Phys. {\bf A 381} (1982) 294
\bibitem{becc} F.D. Becchetti and G.W. Greenlees, Phys. Rev. {\bf 182} (1969) 1190
\bibitem{fil83} B.W. Filippone at al., Phys. Rev. Letts. {\bf 50} (1983) 412
\bibitem{kav69} W. Kavanagh et al., Bull. Amer. Soc. {\bf 14} (1969) 1209
\endlist

\end{document}